# A Comparison of Continuous and Stochastic Methods for Modeling Rain Drop Growth in Clouds

Rehan Siddiqui[1], and Brendan M. Quine[1,2]

[1]Department of Physics & Astronomy, York University, Toronto, Canada

[1,2]Department of Physics & Astronomy & Department of Earth, Space Science and Engineering, Toronto, Canada

Rehan Siddiqui, [1]Department of Physics and Astronomy, York University, 4700 Keele Street, Toronto, Ontario, M3J 1L3, Canada.
E-mail: rehanrul@yorku.ca

Brendan M. Quine, [2]Department of Earth, Space Science and Engineering, York University, 4700 Keele Street, Toronto, Ontario M3J 1L3, Canada.
E-mail: bquine@yorku.ca

# ABSTRACT

Two models for raindrop growth in clouds are developed and compared. A continuous accretion model is solved numerically for drop growth from 20-50 microns, using a polynomial approximation to the collection kernel, and is shown to underestimate growth rates. A Monte Carlo simulation for stochastic growth is also implemented to demonstrate discrete drop growth. The approach models the effect of decreased average time between captures as the drop size increases. It is found that the stochastic model yields a more realistic growth rate, especially for larger drop sizes. It is concluded that the stochastic model showed faster droplet accumulation and hence shorter times for drop growth.

**Key words:** raindrop growth, continuous collection, stochastic collection, Monte Carlo Method, Implicit and semi-implicit technique.

# 1.  Introduction

In a cloud, the development of a size distribution of rain drops with radius R, as they collect droplets of radius $r$, is described by a nonlinear differential equation relating the mean number concentration of droplets $N(r)$ to the rate at which drops and droplets collide and coalesce. The effect of mixing between upwards and downwards moving entities is to reduce the concentration of droplets in the ascending air. The super supersaturated created in the updraft is then distributed over fewer drops, permitting them to grow to larger sizes.

Rain drop collision does not guarantee coalescence. When a pair of drops collides they may subsequently (i) bounce apart, (ii) coalesce and remain so, (iii) coalesce temporarily but then break apart, retaining their initial identities (iv) coalescence temporarily but then break apart to a number of smaller drops. For sizes smaller than 100 microns in radius, the important interactions are (i) and (ii), described by Barnet (2011) and Rogers et al. (1989).

In stochastic raindrop growth, coalescence can broaden the droplet spectrum, but is hindered in the early growth stages by the fact the collection efficiencies between small droplets are extremely small. Coalescence is not sufficient to account for rain development over short periods as shown by an earlier study Robertson (1973). It is now recognized that statistical effects are crucial in the early stages of coalescence. Consequently a stochastic coalescence model provides a convenient means to describe this process Kostinski et al. (2005).

According to Rogers et al. (1989), as droplets grow, their collection efficiencies increase, increasing the probability of coalescence. Once it begins coalescence proceeds rapidly, as indicated by the fast decline in the number of drops. At the same time, super saturation increases sharply because the drops, now fewer in number, are no longer able to consume the excess vapor at the rate it is created.

In general the continuous and stochastic growth of rain drop are classify by the relative amount of water collected from the different sizes of small droplets to large droplets, which is mainly depends upon the mass and size of the droplets. Droplets growing according to the continuous model collect most of their water by capture of droplets while droplets growing by stochastic model collect water from droplets of all the small sizes. According to Berry (1967), the average rate of mass and size increase of $n^{th}$ droplet due to the capture of $r^{th}$ droplets is equal to the product of the collection kernel (volume swept out per unit time and the mass density function (mass per unit volume per unit size of interval).

The effects of turbulence in a cloud can be modeled by a probabilistic collection kernel where the magnitude of the collection kernel indicates, the importance of turbulence (Berry 1967).

In this work we developed and compared two models for raindrop growth in clouds based on continuous accretion and stochastic technique by using numerical solution and Monte Carlo simulation. It is found that the stochastic model yields a more realistic growth rate, especially for larger drop sizes.

We applied MAPLE 13 for supporting numerical techniques and programming features.

## 2. Theory

Consider a collector (larger) drop of radius R that is falling relative to a field of smaller droplets of radius r. The rate at which the collector collides with the smaller droplets is proportional to the shared *collision volume*, $V_c(R,r)$, which is given by the cross-sectional areas of both the drop and the droplet and their vertical velocities $u(R)$, $u(r)$. Derivation and discussion of Equations can be found in Long (1973, 1974) and Robertson (1974).

$$V_c(R,r) = \pi(R+r)^2\{u(R)-u(r)\} \tag{1}$$

The probability that a collision between a drop and a droplet results in an actual capture (coalescence) is described by the collection efficiency $E(R,r)$. Given that the mean number of droplets within the collision volume is $V_c(R,r)N(r)$, where $N(r)$ is the mean number concentration of droplets, the probability per unit time that a drop captures a droplet is

$$\begin{aligned}P(R,r) &= V_c(R,r)N(r)E(R,r) \\ &= \pi(R+r)^2\{u(R)-u(r)\}N(r)E(R,r)\end{aligned} \tag{2}$$

The realistic growth of a collector drop is discrete, where capture of each droplet increases the mass of the drop $M(R)$ by the finite droplet mass $m(r)$. The collector drop also grows stochastically, where each capture has a probability between 0 and 1. The mean growth rate of the collector drop is described by

$$\frac{dM(R)}{dt} = m(r)P(R,r) \tag{3}$$

As a first approximation, we can consider the simplest type of model for collection growth, the continuous model, as

$$\frac{dM(R)}{dt} = m(r)\pi(R+r)^2\{u(R)-u(r)\}N(r)E(R,r) \quad (4)$$

$$\frac{dM(R)}{dt} = K(R,r)w_L(r) \quad (5)$$

Here we have two factors [2]: the droplet *collection kernel* $K(R,r) = \pi(R+r)^2\{u(R)-u(r)\}E(R,r)$, and the liquid water content of the droplets, $w_L(r) = m(r)N(r)$.

A method for deriving an analytical solution for the droplet collection equation, using a polynomial approximation to the kernel, $K_P(R,r) = cx^2$.

Here c is a scaling factor and $x \equiv V(R)$ is the collector drop volume. Then the collection equation becomes,

$$\frac{dM(R)}{dt} = cV^2 m(r)N(r)$$

$$\frac{dV(R)}{dt} = cV^2 v(r)N(r) \quad (6)$$

Here, *v(r)* is the droplet volume. An analytical solution for *V(t)* is found by integrating the above equation, to give

$$V(t) = \frac{1}{(1/V_0 - cNvt)} \quad (7)$$

where $V_o$ is the initial collector drop volume and $c = 1.1 \times 10^{10}$ $cm^{-3}s^{-1}$ is the constant related to the polynomial kernel according to Long et al. (1974).

## 3. Method

For the continuous model of collection growth, equation (8) is numerically solved using an implicit or semi-implicit integration scheme. The implicit scheme is

$$\frac{(V_{n+1} - V_n)}{\Delta t} = cv(r)N(r).V_{n+1}^2 \tag{8}$$

The semi-implicit equation is

$$\frac{(V_{n+1} - V_n)}{\Delta t} = cv(r)N(r).V_{n+1}V_n \tag{9}$$

A Monte Carlo simulation of stochastic drop growth is also implemented. First we calculate the time interval $\Delta t$ to perform a discrete simulation step for which the probability of capture $q = P(R,r)\Delta t$, where q is chosen to be a small value such as 0.1 as suggested by Long (1973). If a uniformly distributed random number x between 0 and 1 is generated and $x > q$, then no capture occurs during the time interval $\Delta t = q/P(R,r)$. If $x \leq q$, a capture is deemed to have occurred and $M(R)$ is increased by m(r). Before the next time step, $P(R,r)$ and $\Delta t$ are recalculated using the new value of $R$ according to Robertson (1974).

### 3.1 *Droplet Terminal Velocity*

One important factor in drop formation is the droplet terminal velocity. In general when downward net gravitational force is equal to upward drag force (i.e. $F_G = F_{drag}$), the droplet reaches a steady fall speed, its terminal velocity. Terminal velocities depend mainly on the size of the droplet. Fig.1. shows the droplet terminal velocity as a function of its radius, with different droplet regimes showing different behaviors by Rinehart (1990). By Rogers et al. (1989), for small droplet sizes ($r \leq 30\mu m$), flow is completely dominated by air viscosity, and the terminal velocity increases quadratically: $u = k_1 r^2$ with $k_1 = 1.19 \times 10^8 \, s^{-1} m^{-1}$. For larger sizes ($30 \mu m \leq r \leq 10^3 \mu m$), flow is turbulent, and the velocity grows linearly: $u = k_3 r$ with $k_3 = 8 \times 10^3 \, s^{-1}$.

### 3.2  Collection Efficiency

The probability that a collision between a drop of radius $R$ and a droplet results in a capture is called efficiency and is given by $E(R,r) = x_o^2/(R + r)^2$. The value of $R$ is important for any size of collector drop and $E$ is small for small values of $r/R$. The collision efficiency as a function of drop radius R increases with drop size, as shown in Fig. 2.

## 4.  Results

Drop growth was computed for an initial collector drop radius of $R_i = 20 \, \mu m$ and continued until the drop reached a final radius $R_f = 50 \, \mu m$. The collected droplets had a radius of $r = 10 \mu m$ and a concentration $N(r) = 100 \, cm^{-3}$. For continuous growth both numerical techniques were applied and the results are plotted in Fig. 3. along with the analytical solution. The stochastic growth was computed using a capture probability of q=0.1. The average growth time by using Monte Carlo runs are also shown in Fig. 3. While the average result shows the continuous growth curves are in close agreement, it is evident that the drop growth rate becomes slower than the Monte Carlo solution as the drop radius increases.

In the Monte Carlo technique, the average time between captures gets smaller as the drop grows. As expected, after a sufficiently large number of captures i.e. at a larger drop radius R, the growth curves stabilize, and increase in parallel to the continuous growth curve also explained by Robertson (1974). The various Monte Carlo runs exhibit statistical variations, but yield shorter average growth times than the continuous model, since their rates increase substantially once the collector drop radius exceeds about 25 microns.

4.1    Model Sensitivity

To explore the statistical behavior and accuracy of the discrete model, a large number (N=1000) of Monte Carlo runs were performed, yielding a distribution of drop growth times, shown in Fig. 4. This distribution has a mean growth time, $T_{avg}$ = 4445 s, with a standard deviation $\sigma$ = 953 s (a 22 % uncertainty)

To check the sensitivity of the model to the capture probability, average growth-times $T_{avg}$ (q) were computed for 100 values of q in the range [0.01, 1.0]. The resulting values are shown in Fig. 5., and their distribution is shown in Fig. 6., with a mean $< T_{avg} >$ = 4434 s and $\sigma$ = 32 s. This demonstrates the low sensitivity of the model to variation of q, with only 0.7% variation in the average growth time.

# 5.    Conclusions

Continuous and stochastic models have been used to simulate the accretion growth of an individual collector drop from a starting size of 20 microns to a final size of 50 microns. In the continuous accretion case, the time for drop growth is unrealistically long due to large accumulation of water contents. In contrast, the stochastic model showed faster droplet accumulation and hence shorter times for drop growth. For a fixed choice of capture probability q=0.1, the average growth time $T_{avg}$ has an uncertainty of 22%. However the sensitivity of $T_{avg}$ to the capture probability was found to be small:

when q is varied between 0.01 and 1.0, it showed only a 0.7% variation. Finally, it is concluded that all the water mass moves with the mode in the stochastic model, whereas in the continuous model, most of the water mass must remain on the small droplets. This work can play a significant role for the analysis for any future rain drop development methodology and any theoretical numerical weather forecasting test.

## *Acknowledgements.*

This work is supported by Physics & Astronomy and Earth and Space Science Department of York University. The author's wishes also to acknowledge Prof. Marry Ann Jenkins of York University and members of Thoth Technology Inc. for their kind technical support concerning various aspects of the material presented here.

## Caption:

**Fig. 1.** Droplet terminal velocity as a function of droplet size 'r'

**Fig. 2.** Collection efficiency as a function of drop radius R, for collisions with droplets of radius r = 10μm. The values are taken from Rogers and Yau, 1989.

**Fig. 3.** Collector drop radius R as a function of time for continuous and stochastic growth models. The analytical solution is shown as a solid red line, with semi-implicit and implicit numerical solutions shown as circles and squares, respectively.

**Fig. 4.** Distribution of collector drop growth times T, obtained from N=1000 Monte Carlo trials with q=0.1: $T_{avg}$ = 4445 s, $\sigma$ = 953 s.

**Fig. 5.** Distribution of average growth times obtained using 100 equally-spaced values of q in the range from 0.01 to 1.0: $<T_{avg}>$ = 4434 s, $\sigma$ = 32 s

**Fig. 6.** Distribution of average growth times obtained using 100 equally-spaced values of q in the range from 0.01 to 1.0.

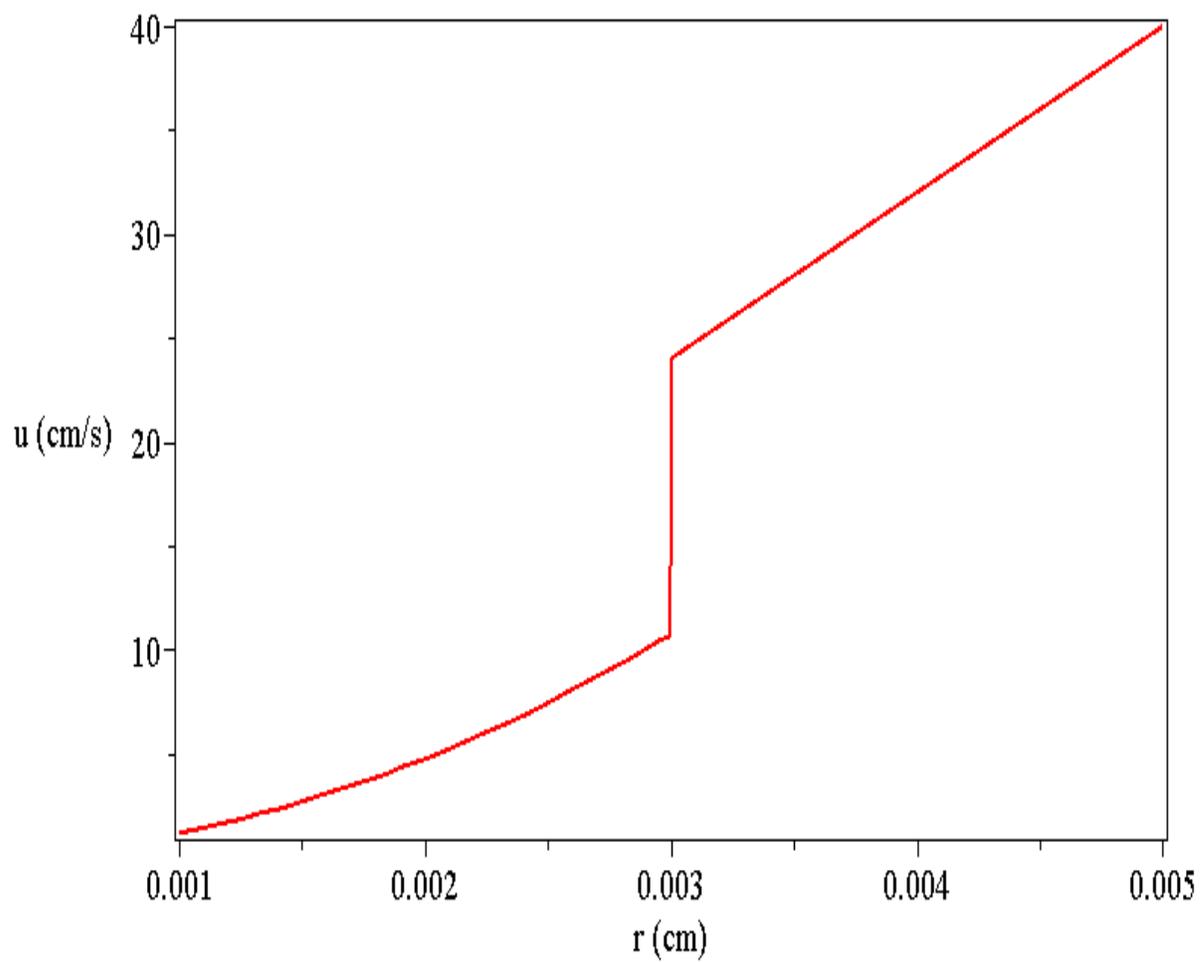

**Fig. 1.** Droplet terminal velocity as a function of droplet size 'r'

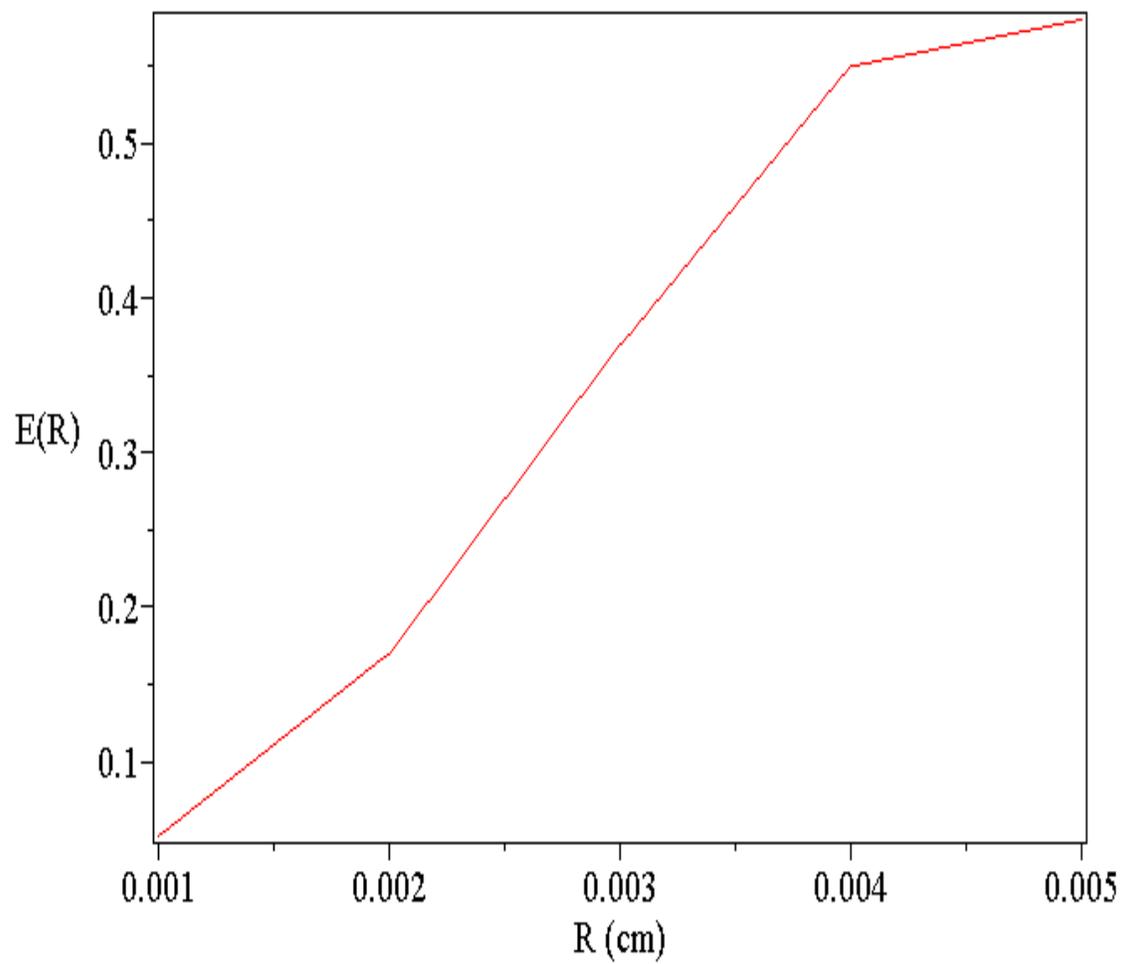

**Fig. 2.** Collection efficiency as a function of drop radius R, for collisions with droplets of radius r = 10µm. The values are taken from Rogers and Yau, 1989.

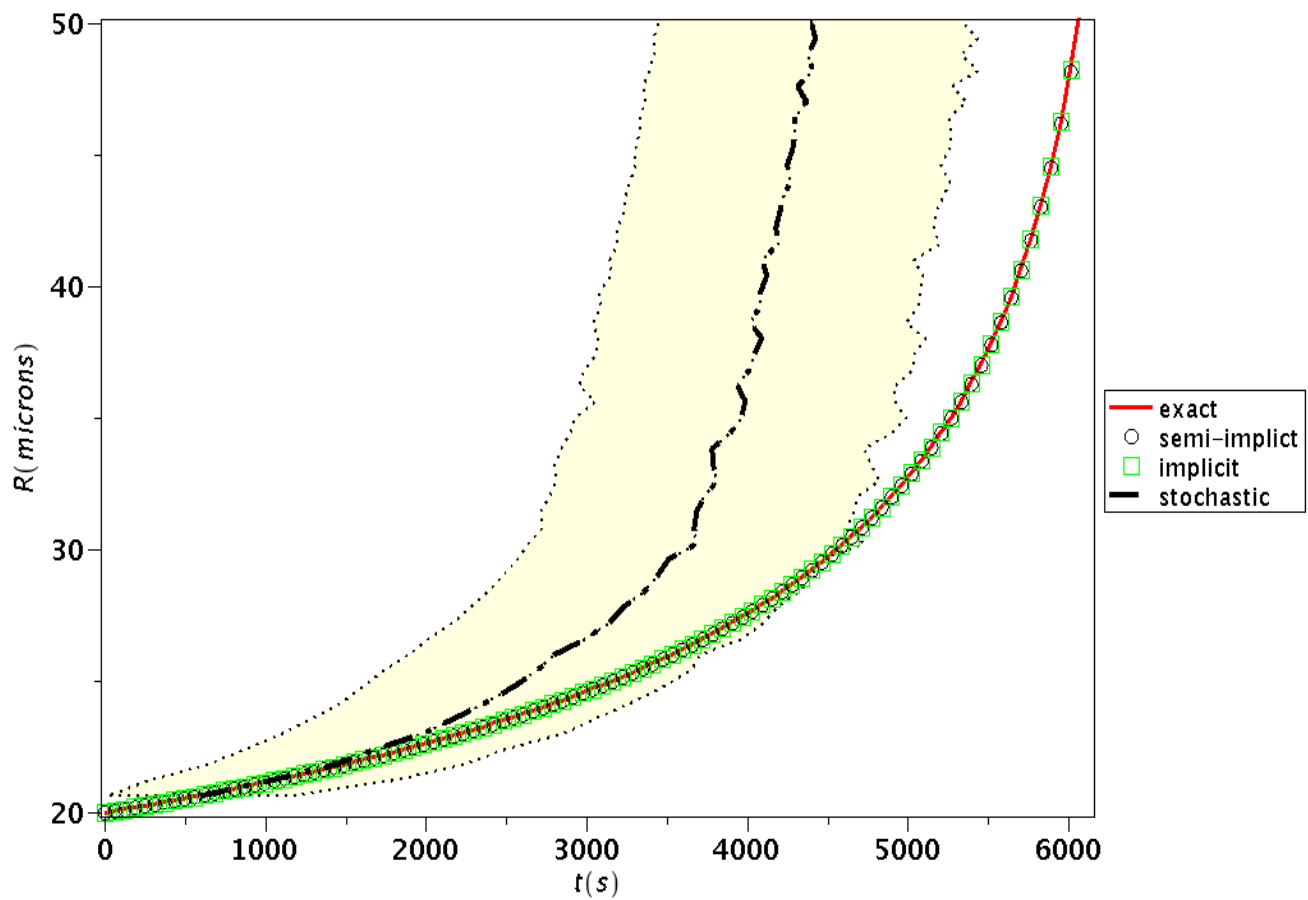

**Fig. 3.** Collector drop radius R as a function of time for continuous and stochastic growth models. The analytical solution is shown as a thick red line, with semi-implicit and implicit numerical solutions shown as circles and squares, respectively. The average growth time computed with the stochastic model is plotted as a thick dashed-dot line, with the two standard deviation range bounded by the dotted lines and shaded in yellow.

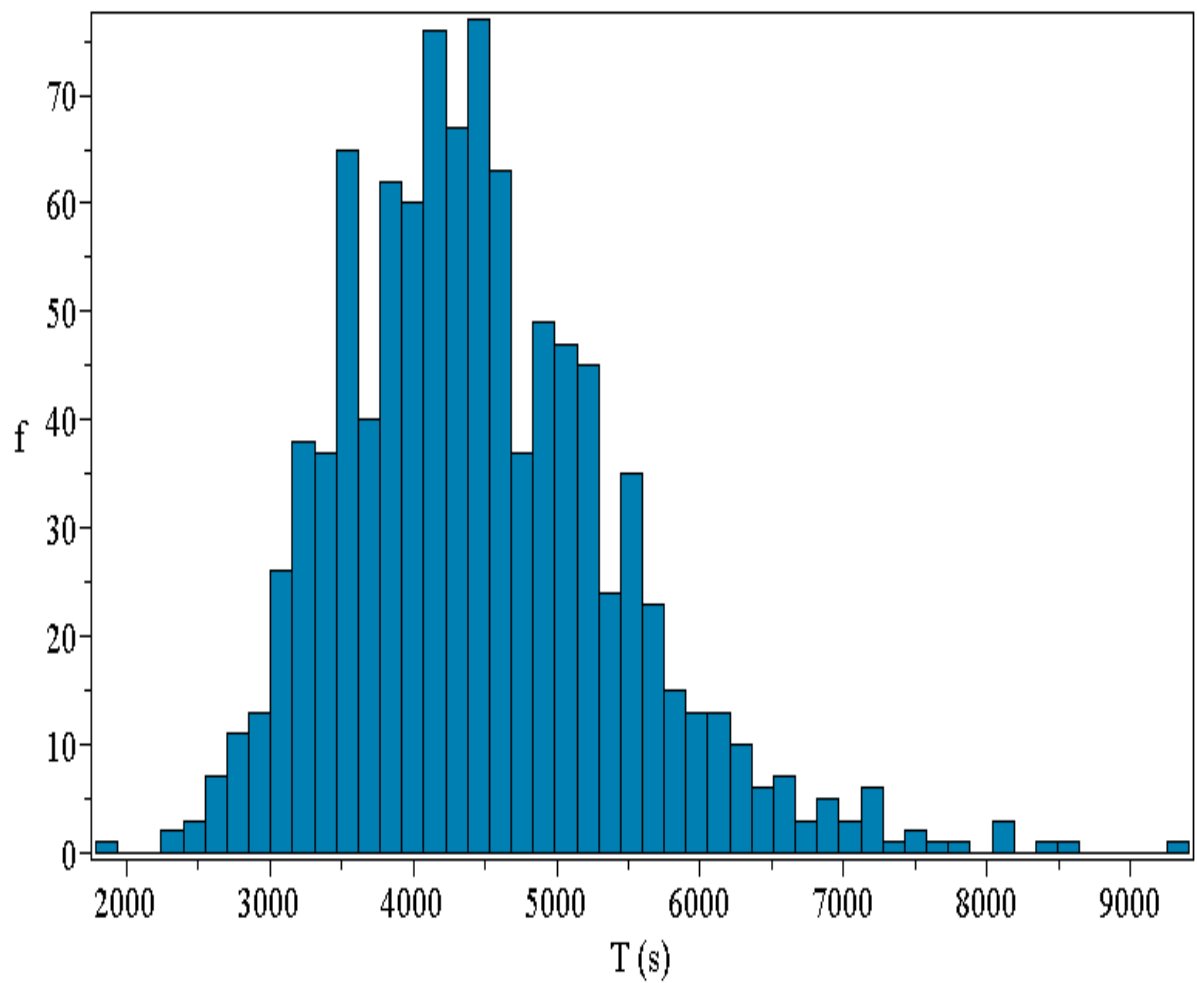

**Fig. 4.** Distribution of collector drop growth times T, obtained from N=1000 Monte Carlo trials with q=0.1: $T_{avg}$ = 4445 s, σ = 953 s.

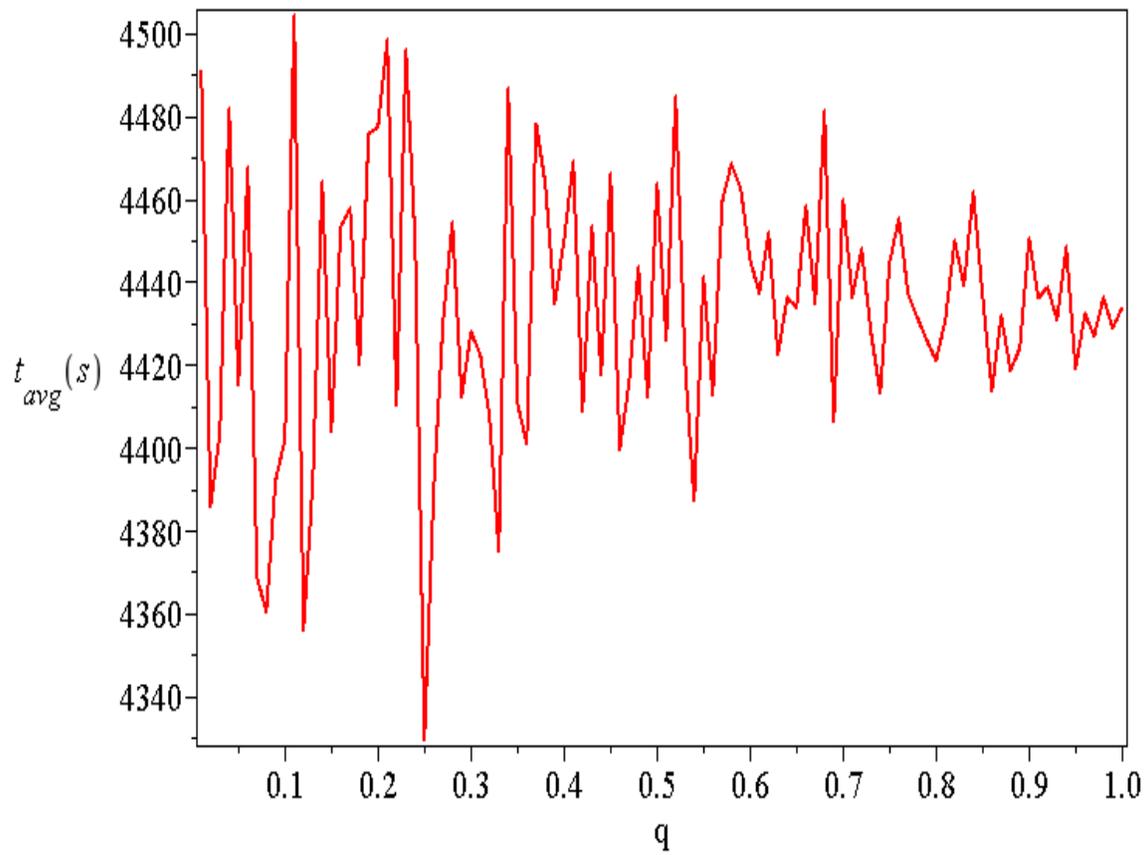

**Fig. 5.** Average growth times obtained using 100 equally-spaced values of q in the range from 0.01 to 1.0.

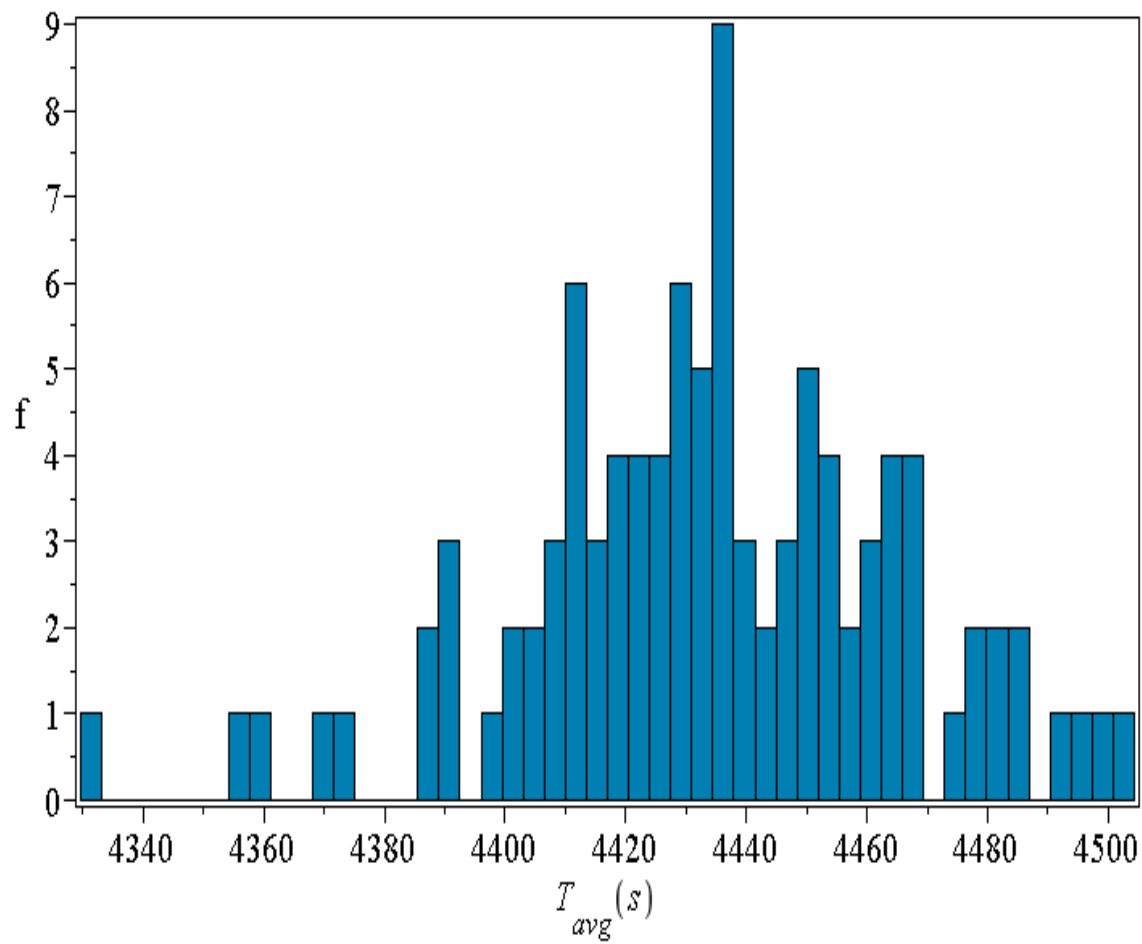

**Fig. 6.** Distribution of average growth times obtained using 100 equally-spaced values of q in the range from 0.01 to 1.0: < $T_{avg}$ > = 4434 s, σ = 32 s